\documentclass{ws-ijmpa}
\usepackage{floatflt}

\newlength{\ziffer}

\newcommand{\TeV}{\,\mbox{Te\kern-0.2exV}}
\newcommand{\GeV}{\,\mbox{Ge\kern-0.2exV}}
\newcommand{\mGeV}{\,\mathrm{Ge\kern-0.2exV}}
\newcommand{\MeV}{\,\mbox{Me\kern-0.2exV}}
\newcommand{\keV}{\,\mbox{ke\kern-0.2exV}}
\newcommand{\eV}{\,\mbox{e\kern-0.2exV}}

\newcommand{\ipb}{\,\mbox{pb}^{-1}}

\newcommand{\bea}{\pagebreak[3]\begin{samepage}\begin{eqnarray}}
\newcommand{\eea}{\end{eqnarray}\end{samepage}\pagebreak[3]}
\newcommand{\beq}{\begin{equation}}
\newcommand{\eeq}{\end{equation}}

\newcommand{\eq}[1]{Eq.~(\ref{#1})}

\begin{document}

\markboth{Daniel Wicke}
{Top Pair Production Cross-Section in the All-Hadronic Channel}

%
\catchline{}{}{}{}{}
%

\title{TOP PAIR PRODUCTION CROSS-SECTION MEASUREMENT IN THE ALL-HADRONIC CHANNEL AT CDF AND D\O}

\author{\footnotesize DANIEL WICKE}

\address{Fermilab, P.O. Box 500\\
Batavia, Illinois 60510,
USA\footnote{on leave of absence from Bergische Universit\"at Wuppertal}\\[1.4mm]
\em on behalf of the CDF and the D\O\ Collaborations
}

\maketitle

\pub{
~}{~}

\begin{abstract}
Measurements of the {$t\bar t$} production cross-section at 
$\sqrt{s} = 1.96$~TeV in proton-antiproton collisions were performed by the CDF and
D\O\ collaborations using $t\bar t$ final states where both $W$'s decay
hadronically. Each experiment uses luminosity of about 160~pb$^{-1}$ of data
collected during Run II of the Fermilab Tevatron collider. Beside kinematical
observables both experiments take advantage of identifying $b$~jets based on
lifetime tagging. While CDF counts the number of $b$-tags over background
after a cut based analysis, D\O\ counts the excess of events after a neural
network based kinematical selection and requiring a $b$-tagged jet.

CDF obtains
$\sigma_{t\bar t}=7.8\pm2.5_\mathrm{stat}~^{+4.7}_{-2.3} \,_\mathrm{syst}$;
~~D\O\,
$\sigma_{t\bar t}=7.7~^{+3.4}_{-3.3}\,_\mathrm{stat}~^{+4.7}_{-3.8} \,_\mathrm{syst}\pm0.5 \,_\mathrm{lumi}$.
\keywords{top; cross-section; Tevatron.}
\end{abstract}

\section{Introduction}
The  measurement of the $t\bar t$  cross-section is an important cross check of our
understanding of the top-production mechanism and of the completeness of its
decay modes. Such measurements are performed  in the ``dilepton'', the
``lepton+jets'' and the ``all hadronic'' decay channel, which are named after 
the decays of the $W$'s, because in the Standard Model the top almost always decays to a
$b$ and a $W$. 

This note summarises the measurements done in the ``all hadronic'' channel 
by CDF\cite{cdfnote7075} and D\O\cite{d0note4428} 
in $p\bar p$-collisions at $\sqrt{s}=1.96\TeV$.
This channel has the highest branching ratio
of $44\%$, however, it suffers from an
overwhelming background from light quark pair-production which has many orders
of magnitude higher cross-section. The additional jets faking the hadronically
decaying $W$'s stem from hard gluon radiation and splitting.

Following the pieces of the master formula for cross-section measurements
\begin{equation}
\sigma_{t\bar t}=\frac{N-B}{\varepsilon  {\cal L}\cdot\mathrm{BR}}
\label{eq:xsec_master}
\end{equation}
first the selection of signal events in this environment,
then the background and the efficiency determination are discussed.
The branching ratio, BR, is taken from standard model predictions\cite{Berger:1997gz,Bonciani:1998vc,Kidonakis:2003qe}.
Luminosity measurements for CDF and D\O\ are discussed elsewhere\cite{Acosta:2001zu,d0lumi}.

\section{Signal Selection}
\subsection{Trigger and Preselection}
At $p\bar p$ colliders signal selection begins with the trigger. 
To select all hadronic $t\bar t$ events CDF and D\O\ require 4 clusters
with a transverse energy above $15\GeV$ and $12\GeV$, respectively. 
CDF in addition requires a minimal
value for the scalar sum of all transverse momenta $H_T\ge125\GeV$, while D\O\
asks the two hardest jets to have $p_T>25\GeV$ and the third jet 
$p_T>15\GeV$.

Both experiments confirm their trigger selection with offline objects and 
veto on isolated high $p_T$ leptons to keep the analyses orthogonal to
their measurements in the other channels.
While D\O\ continues with 6 or more jet
events CDF keeps the 4 jet events for calibration in the background
determination.

Both experiments work with an integrated luminosity of around $160\ipb$.

\subsection{Final Signal Selection}
General kinematic observables can distinguish between $t\bar t$-signal
and multi-jet background. Top events are expected on average to
have a larger overall energy scale and harder sub-leading jets. Their shape is
expected to be more spherical and less planar. Their rapidity
distribution is more central than that of light quark production.

\subsubsection{Kinematic Observables}
To select signal events CDF uses 
the sum of all transverse momenta, $H_T$, the sum of the transverse momenta
without the two leading jets, $H_T^{\mathrm{3j}}$, Aplanarity and Centrality
for the signal selection. 

D\O\ uses in addition  $\sqrt{s}$, the geometric
mean of the 5th and 6th jets $E_T$, $E_{T_{5,6}}$, the $E_T$ weighted number
of jets, $N^A_{\mathrm{jets}}$,
Sphericity and a momentum weighted rapidity moment, $\left<\eta^2\right>$.
D\O\ also exploits observables obtained from assuming a $t\bar t$-event and
reconstructing the decay chain. The consistency of the reconstructed $W$-
and $t$-masses with the known values, the invariant $t \bar t$-mass, the
invariant $WW$-mass as well as the highest and the second highest dijet-mass 
are considered.

\subsubsection{$B$-Tagging}
Besides the kinematic properties the presence of $b$-jets in $t\bar t$-events 
is an important feature that is used in the signal selection.
Both CDF and D\O\ are equipped with silicon vertex detectors which are used to 
reconstruct displaced secondary vertices stemming from $B$-decays. 
Such secondary vertices are reconstructed using tracks belonging to a jet of
interest. The significance of the displacement of the secondary vertex
from the primary vertex is used to decide whether the jet originates from a $b$-quark.
CDF requires a significance of 3, D\O\ one of 7 to call a jets \em $b$-tagged\em.

\subsubsection{Cut Procedures}
To have optimal cuts in the presence of correlated observables D\O\ uses a
series of neural networks (NN) for the selection. A first neural net (NN0) uses a subset of
kinematic information to perform a final step of loose preselection.
Then one $b$-tag is required. The next NN combines all kinematic observables 
into a single number which is then fed into the final neural net (NN2) which includes the
reconstructed top-properties. After cutting on the output of NN2 D\O\
\em selects 220 events\em.

CDF uses straight cuts in which a linear combination of Aplanarity and
$H_T^{\mathrm{3j}}$  is used to find an optimal cut for these correlated
observables. After the kinematic cuts CDF \em counts 326 $b$-tags \em.

\section{Background Estimation}
Both experiments estimate their background contribution directly from data
by removing the actual $b$-tag and instead weighting the events with the probability
of a random event to be $b$-tagged, the so called Tag Rate Function (TRF).
To provide a background estimate, the TRFs must be measured on background samples.

CDF uses the 4-jet events after the preselection to measure their TRF
as function of $p_T$, $\eta$, number of tracks at the primary vertex, total
number of tracks and Aplanarity.
D\O\ uses all preselected events (after NN0) and parametrises the TRF as
function of $p_T$ and $\eta$ in 4 bins of $H_T$.

The validity of the TRFs is checked by comparing distributions obtained by
applying the TRF on the data with those obtained using the actual $b$-tag.

From these studies CDF expects $264.7\pm17.2$ background \em tags \em
thus claiming 61 surplus $b$-tags.
D\O\ expects $186\pm14.8$ background events, i.e. sees an excess of 34 events.
The uncertainties are obtained by repeating the analysis with TRFs obtained
from samples with varied selection cuts.

\section{Efficiency Determination}
The efficiency for the $t\bar t$ signal to survive the applied cuts is obtained from
$t\bar t$ simulation. The obtained total efficiencies are $ 0.047\pm0.010$ and
$0.058\pm0.001_\mathrm{stat}\pm 0.018_\mathrm{syst}$ for CDF and D\O,
respectively.
The CDF efficiency includes a factor of two needed to account for the expected
two $b$-tags in an ideal experiment when applying \eq{eq:xsec_master}.
CDF also neglected their trigger efficiency, which is believed 
to be above $95\%$.

The two  experiments investigated several systematic uncertainties. In both
cases the by far dominating systematic uncertainty comes from the jet energy
scale resolution. It accounts for $28\%$ uncertainty on the efficiency.
Other uncertainties investigated are below $10\%$.
\section{Summary and Results}
CDF and D\O\ preformed  a measurement of the $t\bar t$ cross-section in the
all-hadronic channel.
Combining the described background estimates from data 
and efficiencies obtained from simulation 
with the luminosity of each experiment the following results are obtained:
\beq
\mbox{CDF:}\;\sigma_{t\bar t}=7.8\pm2.5_\mathrm{stat}~^{+4.7}_{-2.3} \,_\mathrm{syst}\quad
\mbox{D\O:}\;\sigma_{t\bar t}=7.7~^{+3.4}_{-3.3}\,_\mathrm{stat}~^{+4.7}_{-3.8} \,_\mathrm{syst}\pm0.5 \,_\mathrm{lumi}
\eeq
The dominating systematic uncertainties in the final result are the jet energy
scale uncertainty of the selection efficiency and the uncertainty of the
TRF-based background prediction, which gets inflated due to the subtraction 
in  \eq{eq:xsec_master}.
\begin{figure}[b]
  \centering
\epsfig{file=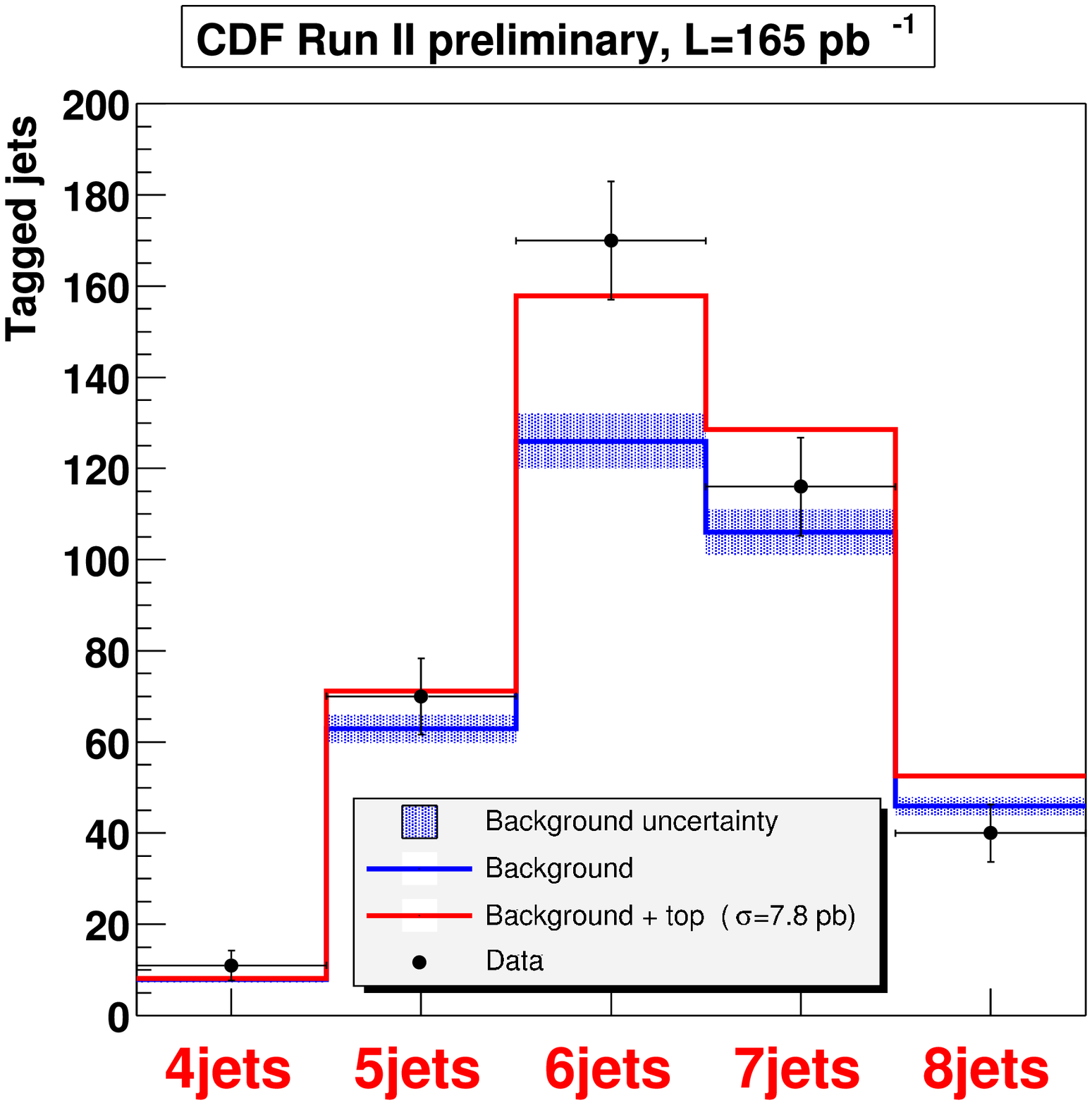,height=.45\textwidth}
\hfill \epsfig{file=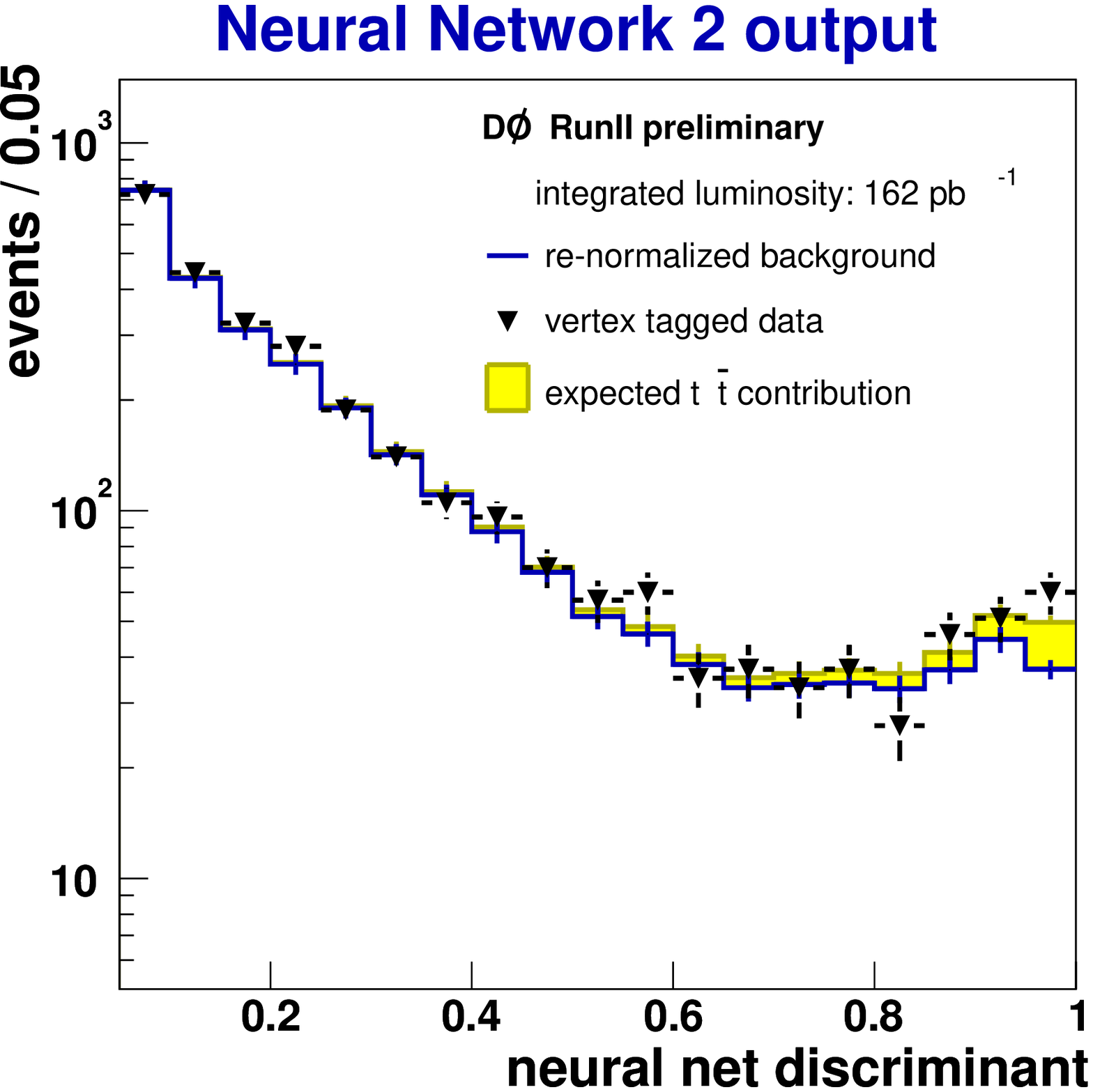,height=.45\textwidth}
\caption{\label{fig:results}%
CDF (left): Number of $b$-tags vs. number of jets per event. Signal is at 6 and above.
D\O (right): The final neural network (NN2) output (required to be above 0.75 for signal).
Background estimates and a Standard Model expectation are
shown by both experiments.}
\end{figure}
\bibliographystyle{unsrtnew}
\begin{flushleft}
\bibliography{Alljets}

\begin{thebibliography}{1}

\bibitem{cdfnote7075}
{\em The~CDF Collaboration}.
\newblock Measurement of the $t\bar t$ production cross section in the
  all-hadronic channel.
\newblock CDF note 7075, August 2004.

\bibitem{d0note4428}
{\em The~D\O\ Collaboration}.
\newblock Measurement of the $t\bar t$ cross section in the all-jets channel.
\newblock D\O\ note 4428, April 2004.

\bibitem{Berger:1997gz}
{\em Edmond~L. Berger and Harry Contopanagos}.
\newblock Threshold resummation of the total cross section for heavy quark
  production in hadronic collisions.
\newblock {\em Phys. Rev.} {\bf D57}(1998)  253--264.

\bibitem{Bonciani:1998vc}
{\em Roberto Bonciani et~al.}
\newblock {NLL} resummation of the heavy-quark hadroproduction cross- section.
\newblock {\em Nucl. Phys.} {\bf B529}(1998)  424--450.

\bibitem{Kidonakis:2003qe}
{\em Nikolaos Kidonakis and Ramona Vogt}.
\newblock Next-to-next-to-leading order soft-gluon corrections in top quark
  hadroproduction.
\newblock {\em Phys. Rev.} {\bf D68}(2003)  114014.

\bibitem{Acosta:2001zu}
CDF Collaboration, {\em D.~Acosta et~al.}
\newblock The {CDF} {C}herenkov luminosity monitor.
\newblock {\em Nucl. Instrum. Meth.} {\bf A461}(2001)  540--544.

\bibitem{d0lumi}
{\em T.~Edwards et~al.}
\newblock Determination of the effective inelastic {$p\bar p$} cross-section
  for the {D\O} {R}un {II} luminosity measurement.
\newblock FERMILAB-TM-2278-E, 2004.

\end{thebibliography}
\end{flushleft}
\end{document}